\documentstyle[useAMS,psfig,epsf]{mn2e}

\newif\ifAMStwofonts
\newcommand{\umin}{u_{\rm min}}
\newcommand{\that}{t_{\rm E}}
\newcommand{\tmax}{t_{\rm max}}
\newcommand{\fr}{F_{\rm r}}
\newcommand{\fb}{F_{\rm b}}
\newcommand{\fu}{F_{\rm u}}
\newcommand{\dol}{D_{\rm OL}}
\newcommand{\dls}{D_{\rm LS}}
\newcommand{\dos}{D_{\rm OS}}
\newcommand{\re}{R_{\rm E}}

\ifCUPmtlplainloaded \else
  \ifAMStwofonts \else % If no AMS fonts

  \fi
\fi

\title{Study by MOA of extra-solar planets in gravitational microlensing events of high magnification}
\author[I.A. Bond et al.]{I.A.~Bond,$^{1,2}$ N.J.~Rattenbury,$^1$ J.~Skuljan,$^2$ F.~Abe,$^3$ R.J.~Dodd,$^{1,4,5}$ 
\newauthor J.B.~Hearnshaw,$^2$ M.~Honda,$^6$ J.~Jugaku,$^7$ P.M.~Kilmartin,$^{1,2}$ A. Marles,$^1$ 
\newauthor K.~Masuda,$^3$ Y.~Matsubara,$^3$ Y.~Muraki,$^3$ T.~Nakamura,$^8$ G.~Nankivell,$^5$ 
\newauthor S.~Noda,$^3$ C.~Noguchi,$^3$ K.~Ohnishi,$^9$ M.~Reid,$^4$ To.~Saito,$^{10}$ H.~Sato,$^8$ 
\newauthor M.~Sekiguchi,$^6$ D.J.~Sullivan,$^4$ T.~Sumi,$^3$ M.~Takeuti,$^{11}$ Y.~Watase,$^{12}$ 
\newauthor S.~Wilkinson,$^4$ R.~Yamada,$^3$ T.~Yanagisawa$^3$, and P.C.M.~Yock$^1$
\\ 
$^1$Faculty of Science, University of Auckland, Auckland, New Zealand\\ 
$^2$Department of Physics and Astronomy, University of Canterbury, Christchurch, New Zealand\\ 
$^3$Solar-Terrestrial Environment Laboratory, Nagoya University, Nagoya 464, 
Japan\\ 
$^4$School of Chemical and Physical Sciences, Victoria University, Wellington, New Zealand\\ $^5$Carter Observatory, P.O. Box 2909, Wellington, New Zealand\\ 
$^6$Institute of Cosmic Ray Research, University of Tokyo, Tanashi, Tokyo 188, Japan\\ 
$^7$Research Institute of Civilization, Tama 206, Japan\\
$^8$Department of Physics, Kyoto University, Kyoto 606, Japan\\
$^9$Nagano National College of Technology, Japan\\ 
$^{10}$Tokyo Metropolitan College of Aeronautics, Tokyo 116, Japan\\ $^{11}$Tohoku University, Sendai, Japan\\
$^{12}$KEK Laboratory, Tsukuba 305, Japan
}
\date{Accepted 0000 January 00.
      Received 0000 December 00}

\pagerange{\pageref{firstpage}--\pageref{lastpage}}
\pubyear{2001}
\begin{document}
\maketitle
\label{firstpage}

\begin{abstract}
A search for extra-solar planets was carried out in three gravitational 
microlensing events of high magnification, MACHO~98--BLG--35, 
MACHO~99--LMC--2, 
and OGLE~00--BUL--12. Photometry was derived from observational images by
the MOA and OGLE groups using an image subtraction technique.
For MACHO~98--BLG--35, additional photometry derived from the MPS and PLANET 
groups was included. Planetary modeling of the three events was
carried out in a super-cluster computing environment.
The estimated probability for explaining the data on MACHO~98--BLG--35 
without a planet is $<1$\%. 
The best planetary model has a planet of
mass $\sim(0.4-1.5)\times M_{\rm Earth}$ at a projected radius of either 
$\sim1.5$ or $\sim2.3$ AU. 
We show how multi-planet models can be applied to the data. We calculated 
exclusion regions for the three events and found that Jupiter-mass planets can 
be excluded with projected radii from as wide as about 30 AU
to as close as around 0.5 AU for MACHO~98--BLG--35 and OGLE~00-BUL-12. 
For MACHO~99--LMC--2, the exclusion region extends out to around 10 AU and 
constitutes the first limit placed on a planetary companion to an
extragalactic star. We 
derive a particularly high peak magnification of $\sim160$ for 
OGLE~00--BUL--12. We discuss the detectability of planets with masses
as low as Mercury in this and similar events.

\end{abstract}


\section{Introduction}

A remarkable diversity of planetary systems has been uncovered in recent 
years through studies of extra-solar planets (Mayor \& Queloz 1995, 
Marcy \& Butler 1998, Perryman 2000). Many further studies will be 
required to obtain a full understanding of this diversity of planetary 
systems and their formation processes. In this paper we report a study 
of extra-solar planetary systems using a gravitational microlensing 
technique. 

\begin{table*}
\centering
\caption{Observational datasets used here. The red and blue passband 
filters used by MOA were 630--1000 nm and 400--630 nm respectively 
(Yanagisawa et al. 2000). We used difference imaging analysis on those 
datasets comprised of raw images. The MPS and PLANET datasets are based on 
DoPHOT analyses previously carried out by the respective groups.}
\label{datasets} 
\begin{tabular}{@{}lllllll}
Event & Telescope & Location & Passband & Number of   & Data Source\\
      &           &          &          & Data Points & \\[10pt]
MACHO~98--BLG--35 & MOA 0.6-m & NZ & red & 162 & Images from MOA-camI\\
        & MPS 1.9-m & Australia & R & 128 & Rhie et al. (2000)\\
        & PLANET 1-m & S. Africa & I & 3 & PLANET website\\
	& PLANET 1-m & S. Africa & V & 5 & PLANET website\\
        & PLANET 1-m & Australia & I & 20 & PLANET website\\
        & PLANET 1-m & chile & I & 18 & PLANET website\\
        & PLANET 1-m & chile & V & 8 &  PLANET website\\
MACHO~99-LMC-2 & MOA 0.6-m & NZ & red & 341 & Images from MOA-camII\\
        & MOA 0.6-m & NZ & blue & 346 & Images from MOA-camII\\
        & OGLE 1.3-m & chile & I & 219 & Images provided by OGLE\\
OGLE~00-BUL-12 & OGLE 1.3-m & chile & I & 300 & Images provided by OGLE \\
\end{tabular}
\label{table1}
\end{table*}   

Mao and Paczynski (1991) pointed out that extra-solar planets could 
be detected using gravitational 
microlensing\footnote{Liebes (1964) may have been the first to consider the
detection of planets by microlensing and also the first to consider
high magnification microlensing events of the type described in this paper}, 
because the characteristic 
scale in gravitational microlensing happens to coincide quite closely 
with the characteristic size of the solar system. The characteristic 
scale of gravitational microlensing is the radius of the Einstein 
ring, $\re$, which is   
\begin{equation}
R_{\rm E} = \sqrt{\frac{4GM_{\rm L}}{c^2}} \times \sqrt{\frac{\dol\dls}{\dos}}.
\label{equation1}
\end{equation}
Here $M_L$ denotes the mass of the lens, and $\dol$, $\dls$ and $\dos$ the 
distances from the observer to lens, lens to source and observer to source 
respectively. For microlensing events that occur in the galactic bulge, 
$M_{\rm L}$ is typically $\sim0.3M_{\odot}$, and $D_{\rm{OL}}$, $D_{\rm{LS}}$ 
and $D_{\rm{OS}}$ are typically about 6 kpc, 2 kpc and 8 kpc respectively. 
These values imply $R_{\rm E}\sim$ 1.9 AU.  Consequently, if other planetary 
systems have the same characteristic size as ours, planets orbiting a lens 
star may perturb the lensing, because light from the source star may pass 
close to one or more of the planets at some time during a microlensing event. 
Gould \& Loeb (1992) and Bolatto \& Falco (1994) showed that 
Jupiter-like planets could be detected with about 20\% efficiency 
in typical microlensing events if they are monitored at about hourly 
intervals with a photometric precision of a few per cent throughout 
each event. The PLANET collaboration has used this technique to set 
an upper limit on the abundance of Jupiter-like planets of order 30\% 
(Albrow et al. 2001). This result is consistent with results obtained 
independently by radial velocity and transit measurements 
(Marcy \& Butler 1998, Charbonneau et al. 2000, Henry et al. 2000, 
Jha et al. 2000). Bennett \& Rhie (2000) emphasized the detectability of
terrestrial planets by microlensing where they showed that Earth-mass
planets could be detected with about 5\% efficiency in typical events if
they are monitored hourly at a photometric precision of better than 1\%.

A modification of the gravitational microlensing plant search strategy
was pointed out by 
Griest \& Safizadeh (1998) in which planets, including those less massive 
than Jupiter, could be detected with high efficiency in microlensing events 
of high magnification. This occurs when the distance of closest approach of 
the lens star to the line-of-sight to the source star, $\umin$, is much 
smaller than the Einstein radius, i.e. when $\umin \ll 1$ where $\umin$ is 
expressed in units of $R_{\rm E}$. In these events a circular, or 
near-circular, image 
generally\footnote{The exceptional situation where ``caustic crossings'' 
occur is discussed in \S5.} 
forms around the lens with radius $R_{\rm E}$ at the time of maximum 
amplification, and the maximum amplification 
$A_{\rm max}\approx 1/\umin \gg 1$. If a planet is orbiting the lens star 
at a projected radius of about a few AU at the time of an event, it will 
perturb the image, although the perturbation will be small because most 
of the image will lie far from the planet and remain unperturbed. For 
planets with masses as low as that of Earth, simulations show 
(see \S5 below) that the perturbation may nevertheless be detectable 
with appropriate observations. The aim of the present study was to 
exploit this sensitivity. 

\begin{figure}
  \centerline{\psfig{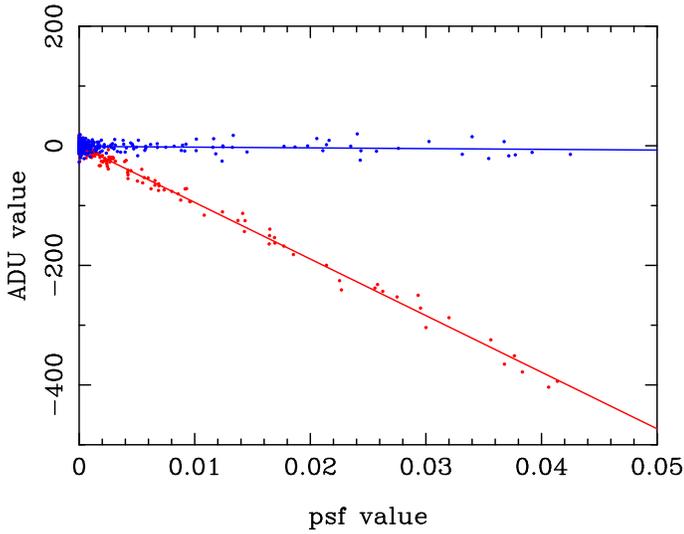}}
  \caption{Fits of empirical PSF profiles to profiles on a subtracted image. 
     The falling fit is for the profile on the first
     subtracted image at the position of MACHO~98--BLG--35. The horizontal 
     fit is for
     one of the vast majority of stars that do not change and are well 
     subtracted.  The slope of any fit provides the $\Delta F$ measurement, 
     and the standard deviation from the fit provides a measurement of the 
     quality of the image subtraction.}
\label{profile}
\end{figure}

\begin{figure}
  \centerline{\psfig{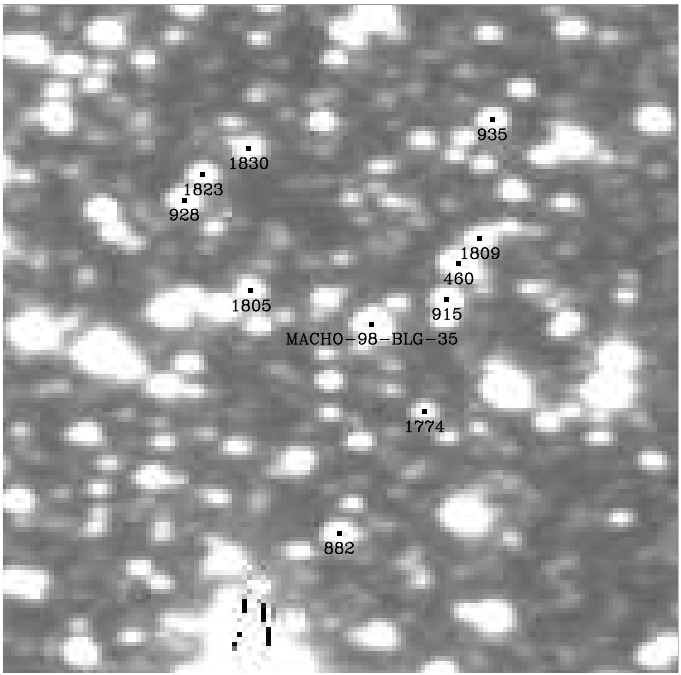}}
  \caption{Sub-region of the reference image used in the difference imaging
     analysis of the MOA dataset on MACHO~98--BLG--35. This image was 
     constructed
     by combining four of the best seeing images which occurred during the 
     times of peak brightness of the event. Also shown are the position
     of ten objects selected as check stars. The camera MOA-cam1, with a 
     pixel size on the sky of 0.65\arcsec, was used for these exposures.}
  \label{finder98}
\end{figure}

\begin{figure}
  \centerline{\psfig{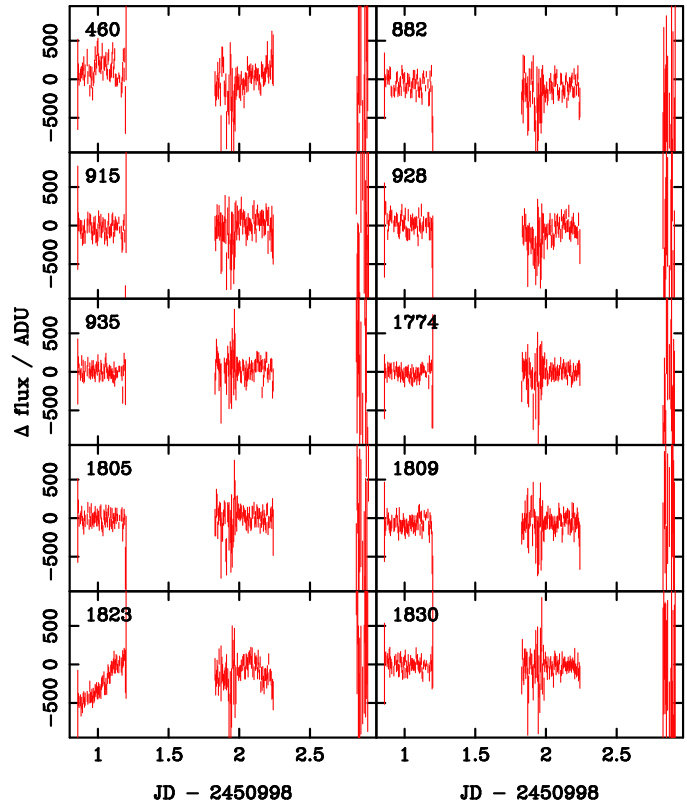}}
  \caption{Light curves of ten check stars for MACHO~98--BLG--35 on the 
  peak night and on the following two nights. Stars 460, 928 and 1823 
  appear to have varied on both nights 1 and 2. The presence of thin 
  cloud during the second quarter of the second night and of thick cloud 
  on the third night is apparent.}
  \label{checks1}
\end{figure}

The first analysis of a high amplification event in terms of planets was 
reported by Rhie et al. (2000). These authors reported a low-mass planetary 
signal (few Earth masses to few Uranus masses) at a marginal level of 
confidence for the event MACHO~98--BLG--35, and excluded Jovian planets 
over an extensive region surrounding the lens star. They also pointed out 
that further such results in additional events would provide statistically 
significant constraints on the abundance of Earth-mass planets, but that 
more accurate planetary parameters can be obtained in events in which a 
"planetary caustic" is crossed or approached, which generally occurs at
lower magnifications.

In this paper we report a re-analysis of 
MACHO~98--BLG--35, where the data in Rhie et al (2000) obtained by MOA
have been analysed using an improved technique based on image subtraction
and the inclusion of additional data. We also report analyses of two more 
events of high magnification, MACHO~99--LMC--2 and OGLE~00--BUL--12. We note 
that the event MACHO~99--LMC--2 coincides with the event OGLE 99--LMC--1 that 
was found independently by the OGLE group. 

\section{Observational Datasets}

Data from the MOA (Abe et al. 1997), OGLE (Udalski et al. 1997), 
MPS (Bennett et al. 1999) and PLANET (Albrow et al. 2001) microlensing 
groups\footnote{MOA website: www.vuw.ac.nz.scps/moa/\\ 
OGLE website: bulge.princeton.edu/\~{}ogle/ \\ 
MPS website: bustard.phys.nd.edu/MPS/ \\ 
PLANET website: thales.astro.rug.nl/\~{}planet/} 
were used in the present analysis. The datasets are summarized in Table~1. 
Wherever possible, we attempted to obtain actual images and analyse 
these using an image subtraction procedure to achieve optimum photometric 
precision in the dense stellar fields in which microlensing is observed. 
The reduced datasets so obtained are posted at the MOA website.

Observations of MACHO~98--BLG--35 and MACHO~99--LMC--2 by MOA were made with
the cameras MOA-cam1 and MOA-cam2 respectively (Yanagisawa et al. 2000).
We note here that MOA-cam2 was operating in a slightly non-linear mode 
during 1999, and that the MOA data on MACHO~99-LMC-2 were corrected for 
this on a pixel-by-pixel basis using linearly calibrated stars from 
observations made in 2000.

Datasets on MACHO~98--BLG--35 by the MPS and PLANET groups were included 
in the present analysis. These datasets were reduced by these groups using 
the DoPHOT procedure of Schechter et al. (1993) and posted at their websites. 
In the case of the MPS dataset, it is the same dataset that was used 
previously by Rhie et al. (2000). In the case of the PLANET dataset, 
the data used here were obtained by extracting the unpublished, graphical 
data at the PLANET website, and allowing for a time offset of 0.007 day 
(A. Gould, private communication, 2000). The accuracy of the graphical
extraction was much better than the error bars on the data. This may be
seen by comparing the PLANET data shown in Fig.~10 of the present paper
(see below) with Fig.~19 of their subsequent publication (Gaudi et al 2002).
The need for the time offset, which
was presumably caused by a heliocentric correction and/or simple clock
errors of a few minutes, became clearly evident from comparisons of
the raw PLANET and raw MPS/MOA datasets for the event
            
\section{Image Subtraction Analysis}

\begin{figure}
  \centerline{\psfig{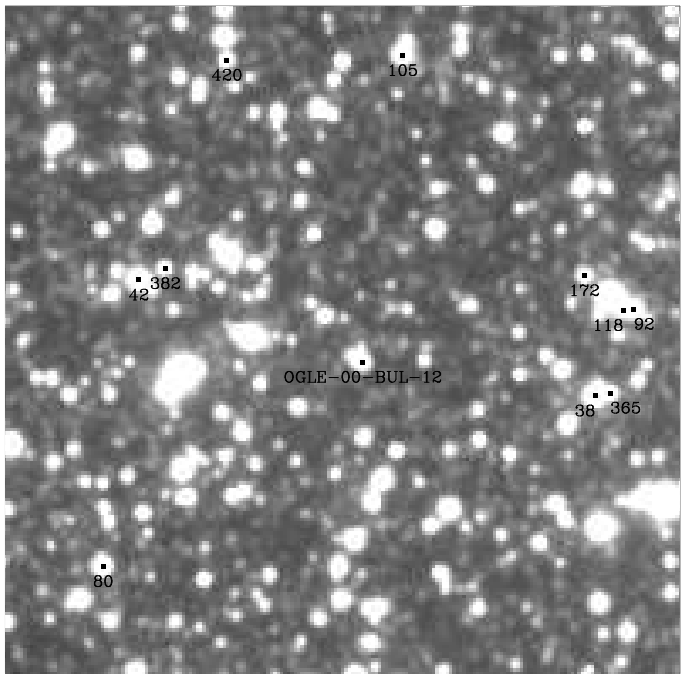}}
  \caption{Sub-region of the reference image used in the difference imaging
     analysis of the OGLE datasets on OGLE~00-BUL-12 together with the 
     positions
     of ten objects selected as check stars. The event itself is strongly 
     blended on the image. Subsequently, a number of blended objects were
     selected as check stars. The plate scale here is 0.42\arcsec per pixel.}
  \label{finderog}
\end{figure}

\begin{figure}
  \centerline{\psfig{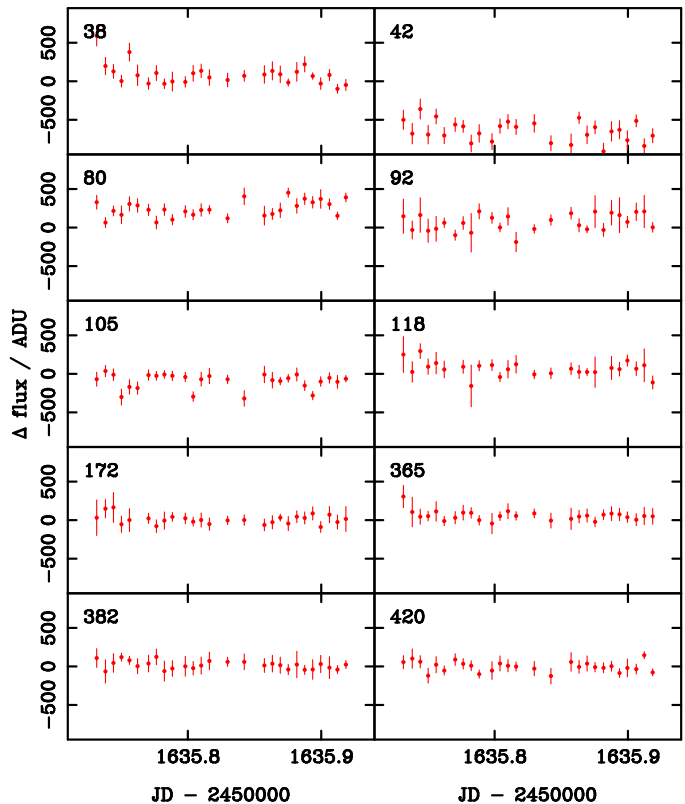}}
  \caption{Light curves of ten check stars for OGLE~00-BUL-12 on the peak 
  night. Stars 38 and 42 appear to have varied.}
 \label{checks2}
\end{figure}

\begin{figure}
  \centerline{\psfig{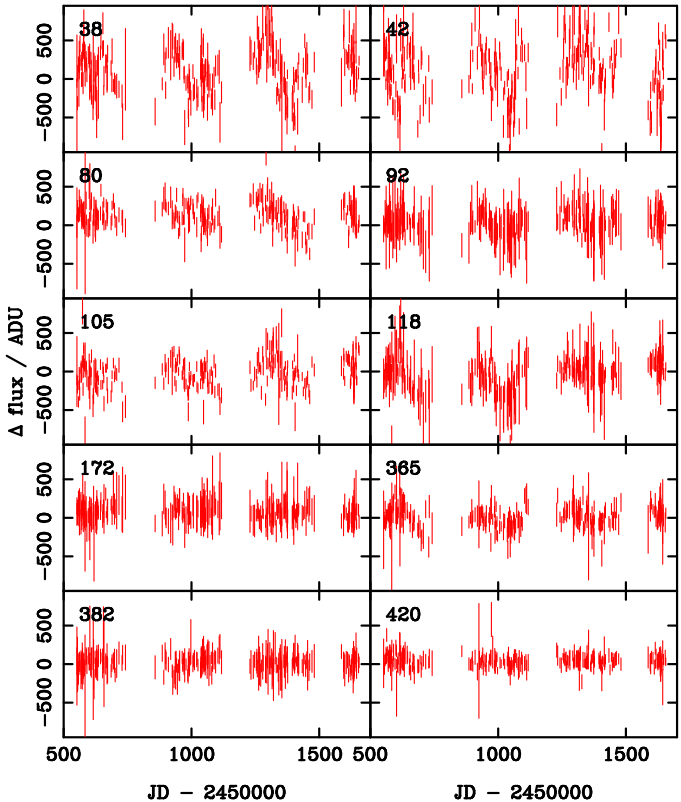}}
  \caption{Light curves of the same stars as in Fig.~4 from 1997 to 2000. 
  The stars 38 and 42 are confirmed as being variable on a time scale of 
  one day.}
 \label{checks3}
\end{figure}

The observations listed in Table~1 by the MOA and OGLE groups were reduced 
by an image subtraction procedure. Data from each telescope in each passband 
were treated as separate datasets. For each dataset, the flatfielded images 
were first geometrically aligned to an astrometric reference chosen from 
amongst the best seeing images. A reference image for the image subtraction 
process was then formed by stacking the very best seeing and signal-to-noise 
images. The convolution kernel which relates the seeing on the reference 
image to that for a particular image was then determined using our own 
implementation of the technique of Alard \& Lupton (1998) for modeling 
the kernel along with the modification of Alard (1999) which models the 
spatial variations of the kernel across the CCD field of view. 

Photometry on the subsequent dataset of subtracted images was carried out 
by first constructing an empirical point-spread-function (PSF) on the 
reference image using 5--6 bright, isolated stars near the event. A PSF 
for a particular image was then formed by convolving the reference PSF 
with the appropriate kernel. This PSF was fitted to the flux profile on 
the subtracted image at the position of a star to obtain a ``delta-flux'' 
measurement, $\Delta F$. The fit was done by re-aligning the centroid of 
the empirical PSF to that of the $\Delta F$ profile on the subtracted image, 
and performing a least squares fit to the pixel by pixel cross-plot of the 
two profiles. Typical fits are shown in Fig.~\ref{profile}. 

\begin{table*}
\centering
\caption{Parameters and statistics of the best single lens fits to the data 
obtained by the MOA and OGLE groups on the three events studied here and 
reduced by image subtraction. The uncertainties given correspond to 
$\Delta\chi^2=6.2$.} 
\label{parameters}
\begin{tabular}{@{}lccccr@{}}

\multicolumn{1}{c}{Event } & 
\multicolumn{1}{c}{$t_{\rm max}$} & 
\multicolumn{1}{c}{$t_{\rm E}$} & 
 $u_{\rm min}$ & $A_{\rm max}$ & 
\multicolumn{1}{c}{$\chi^2$/dof} \\

 & 
\multicolumn{1}{c}{JD$-$2450000} & 
\multicolumn{1}{c}{days} & & \\[10pt]

MACHO~98--BLG--35 & ~999.15  & $27.7\pm^{\infty}_{19.2}$ & $0.0103\pm^{0.0539}_{0.0103}$ & $96\pm^{\infty}_{81}$ & 329.1/157 \\[10pt]
MACHO~99-LMC-2  & 1337.24 & $61.3\pm^{11.4}_{ 9.9}$ & $0.0249\pm^{0.0049}_{0.0041}$ & $40.3\pm^{7.8}_{6.7}$ & 1171.2/897 \\[10pt]
OGLE~00-BUL-12  & 1635.87 & $69.4\pm^{27.6}_{15.4}$ & $0.0063\pm^{0.0019}_{0.0018}$ & $159.4\pm^{65.2}_{36.2}$ & 720.4/295 \\[10pt]

\end{tabular}
\end{table*}

The scatter in fits such as those shown in Fig.~\ref{profile} provides 
an indicator of the quality of the image subtractions. We used the standard 
deviation from the best fit straight line as a measurement of the subtraction 
quality associated with a given $\Delta F$ measurement. In a good quality 
subtraction, the profile of an object on the subtracted image should match 
that of the PSF on the unsubtracted image, and thus have a low standard 
deviation. Uncertainties in the $\Delta F$ measurements for any object 
were determined empirically. The profile fitting technique was applied 
to the positions of all resolved stars in the field. The frame-to-frame 
statistical errors for any particular object were then determined from 
the scatter in the $\Delta F$ values for stars with similar profile standard 
deviations.

Stars close to the microlensing events were selected as checks. Fig.~\ref{finder98} shows stars with similar statistical profile errors as MACHO~98--BLG--35, and Fig.~\ref{checks1} shows their light curves. Figs.~\ref{finderog}--\ref{checks3} show similar stars from the OGLE database on OGLE~00--BUL--12. It is evident from these that the systematic errors in the image subtraction reduction procedure are significantly less than the statistical errors, and that the statistical errors are realistic.

\section{Single Lens Fitting}

\begin{figure*}
  \centerline{\psfig{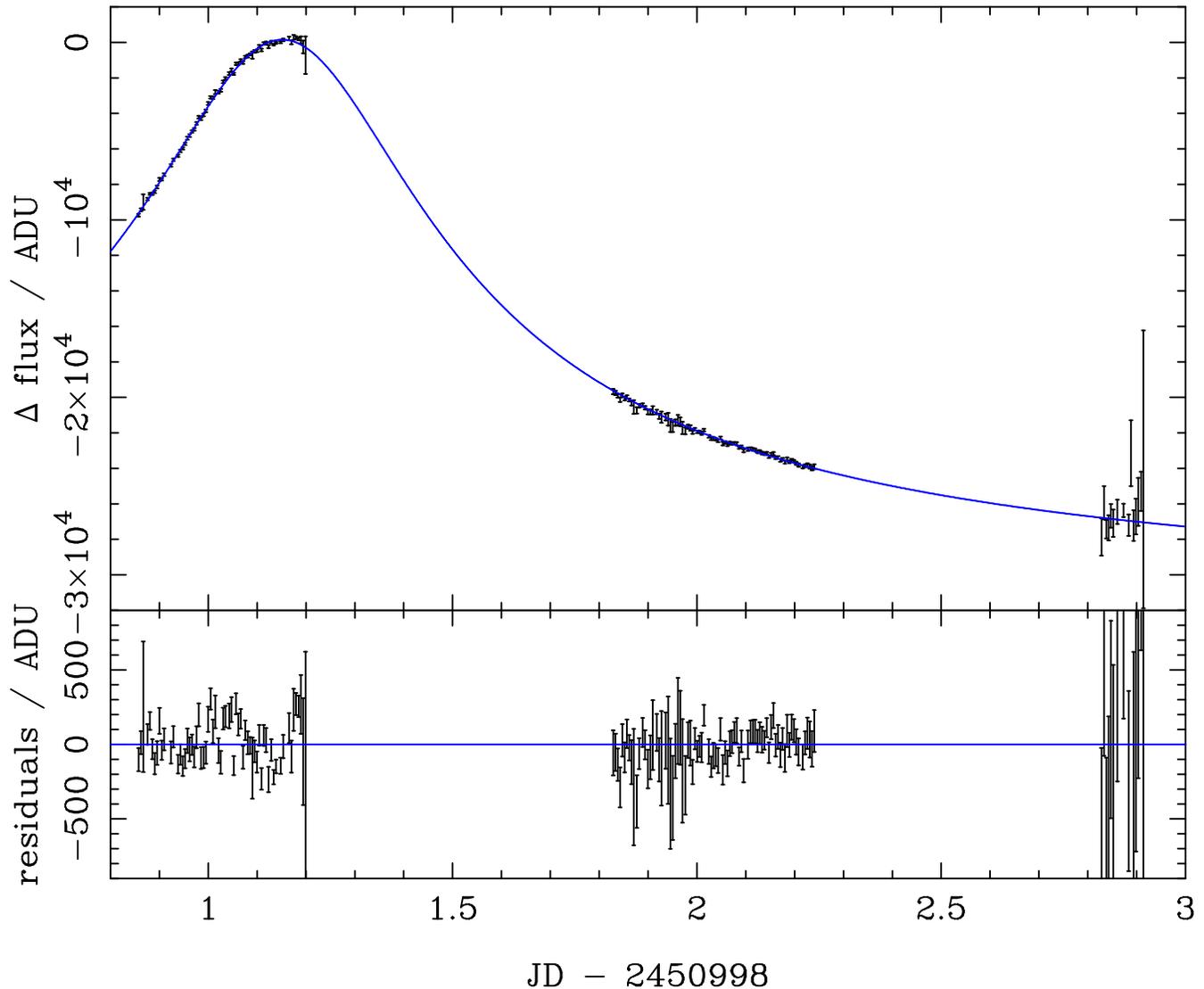}}
  \caption{Light curve for MACHO~98--BLG--35 obtained by the MOA group in 
  the red passband using subtraction photometry. The best single lens fit 
  to the data is also shown. Thick cloud cover occurred on the third night, 
  and thin cloud cover was present during the second quarter of the second 
  night.}
\label{lc98}
\end{figure*}

\begin{figure*}
  \centerline{\psfig{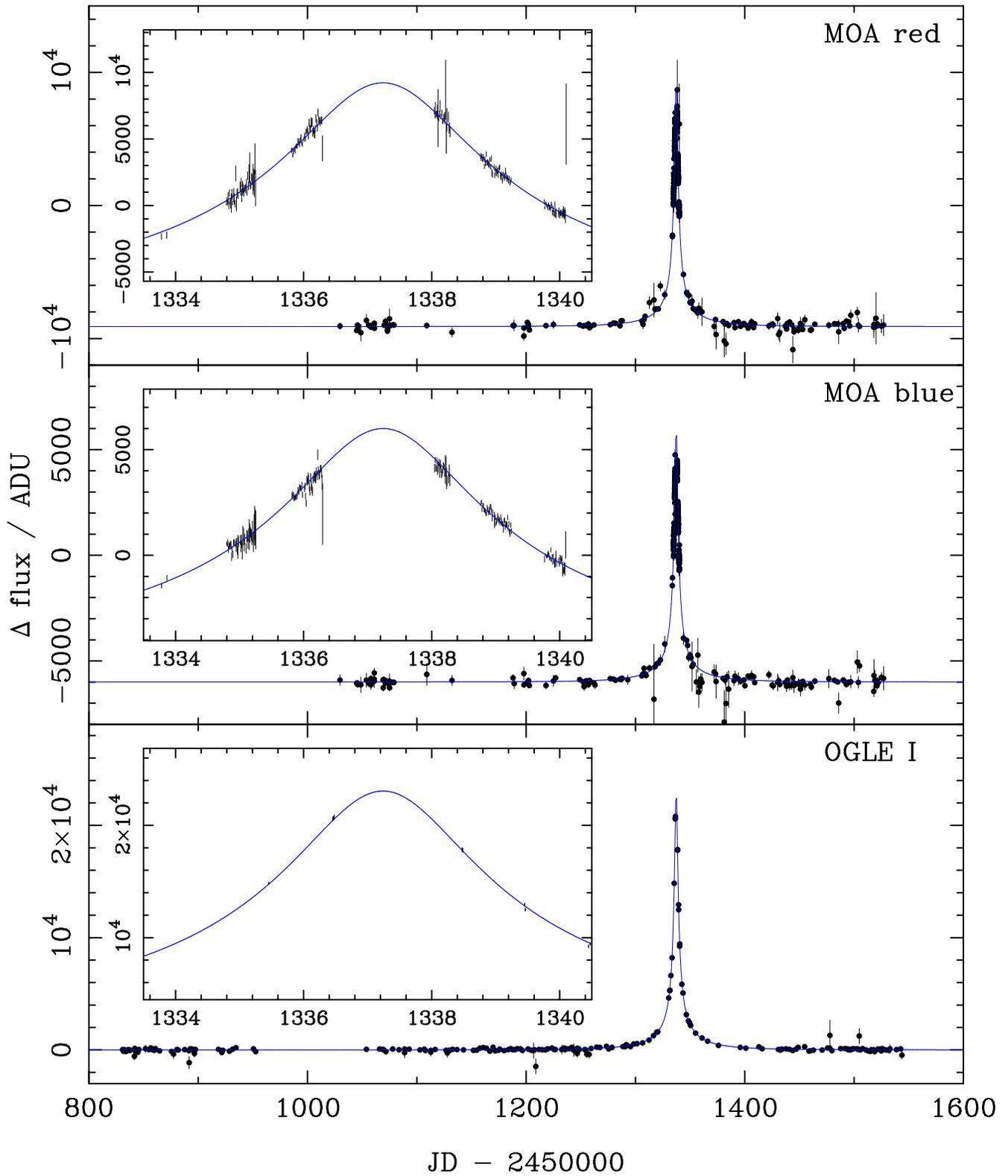}}
  \caption{Light curves in the red, blue and I passbands for MACHO~99-LMC-2 
  on two time scales, and the best single lens fit to the data. The red and 
  blue passband data are by MOA, and the I passband data are by OGLE. All 
  the images were reduced by MOA using subtraction photometry as described 
  in section~3.}
 \label{lc99}
\end{figure*}

\begin{figure*}
  \centerline{\psfig{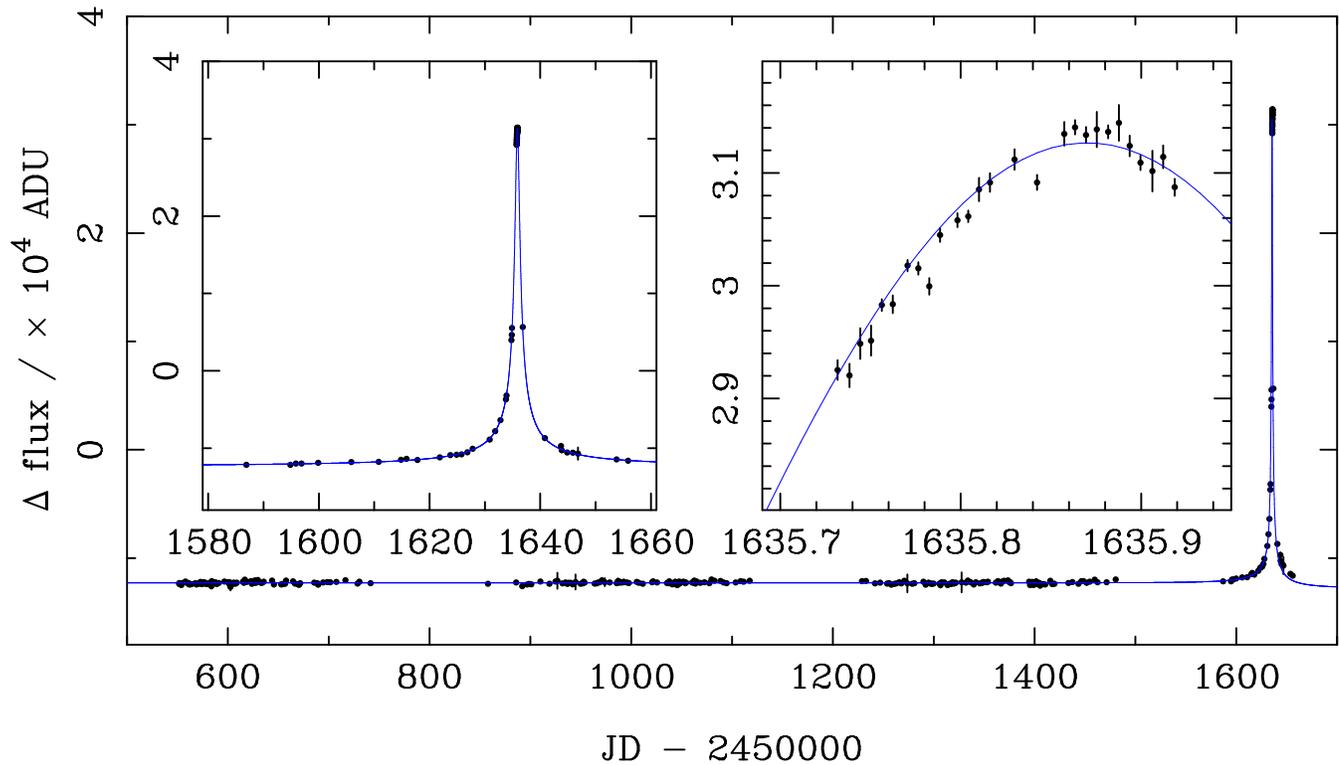}}
  \caption{Light curve in I for OGLE~00-BUL-12 on three time scales, and 
  the best single lens fit to the data. The images were supplied by the 
  OGLE group and reduced by the MOA group using subtraction photometry.}
 \label{lcogle}
\end{figure*}

For the datasets that were obtained by image subtraction the fluxes in any 
passband were fitted to the function 
\begin{equation} 
\Delta F(t) = \fb A(u(t)) - \fr
\label{equation2}
\end{equation}
where $\fb$ denotes the baseline flux of the source star, and $\fr$ is the 
source flux on the reference image used in the image subtraction process, 
which in general contains some lensed 
flux\footnote{For datasets obtained by DoPHOT a similar equation holds, 
viz. $F(t) = \fb A(u(t)) + \fu$, where $\fu$ denotes the flux from nearby 
unresolved stars that are not recognized by the DoPHOT reduction procedure 
as being separate stars, and which are unlensed. This equation was used to 
incorporate the data of the MPS group in the analysis of MACHO~98--BLG--35 
presented in \S5.1.}. 

The well known amplification factor, $A(u)$, expressed in terms of the 
distance $u$ of the lens star from the line-of-sight to the source star 
expressed in units of $R_{\rm E}$, is given by (Paczynski 1986)
\begin{equation}
A(u)=\frac{u^2+2}{u\sqrt{u^2+4}},  
\label{equation3}
\end{equation}
where
\begin{equation} 
u(t)=\sqrt{\umin^2+ {\left(\frac{t-t_{\rm max}}{t_{\rm E}}\right)}^2}.
\label{equation4}
\end{equation}
Here $\umin$ is the minimum value of $u$ during an event and $t_{\rm max}$ 
is the time of maximum amplification, i.e. the time when $u=\umin$. The 
quantity $t_{\rm E}$ is the time scale that characterizes an event. It is 
equal to $R_{\rm E}/v_{\rm T}$ where $v_{\rm T}$ is the transverse velocity 
of the lens with respect to the line-of-sight to the source. In fitting a 
single lens curve to photometry taken in one or more passbands for a given 
event, there are therefore three microlensing parameters and two flux 
parameters for each passband.

The light curves for the three events are shown in 
Figs.~\ref{lc98}--\ref{lcogle} together with 
the best single lens fits to the data. For MACHO~98--BLG--35 only the 
data obtained by the MOA group have been included, since these were the 
only data to be reduced by image subtraction. The light curve for this 
event appears to exhibit substantial deviations from the single lens fit 
near its peak, but not at other times, suggestive of the presence of a 
planet (or planets) orbiting the lens star. Other possible causes of a 
deviation near the peak include the source star being a binary or having
starspots. The binary interpretation seems unlikely (Rhie et al 2000). Also
it would seem that starspots are unlikely to produce perturbations of
the same strength as planetary deviations except in extreme situations
(Rattenbury et al 2002, submitted to MRAS). For MACHO~99-LMC-2 no data 
were available to us at the peak, and in this case no apparent deviation 
from the single lens light curve is seen. For OGLE~00-BUL-12 there appears 
to be a possible deviation near the peak. The parameters and statistics of 
the single lens fits are displayed in Table~2. 

We note that the maximum magnification determined for OGLE~00-BUL-12 by 
image subtraction, $A_{\rm max}=159$, is considerably higher than the 
value of 50 first reported by the OGLE Early Warning System. It should 
be noted that the first analysis was carried out using DoPHOT and that 
there is a significant degree of blending present in this event. This 
would underestimate the fitted value for the peak magnification. On the 
other hand, the light curve presented here is based on image subtraction 
photometry and should be unaffected by blending.

\section{Planetary Modeling}

We used the inverse ray shooting technique of Wambsganss (1997) to simulate 
lensing by planetary systems of arbitrary complexity, by firing photons 
from the telescope through the lens system to the plane of the source star. 
For every component of the lens system a deflection $4GM/bc^2$ was applied, 
where $b$ denotes impact parameter. Photons that hit the source star were 
retained; those that missed were discarded. In this way the finite size of 
the source star was allowed for, but the lens was treated in the thin-lens 
approximation. All the lens components were treated as if they were in a
plane perpendicular to the ray direction. This treatment is expected to be
sufficiently accurate because the dimensions of the lens system are much
smaller than the distances between the lens and the observer, and the
lens and the source. 

\begin{figure}
  \centerline{\psfig{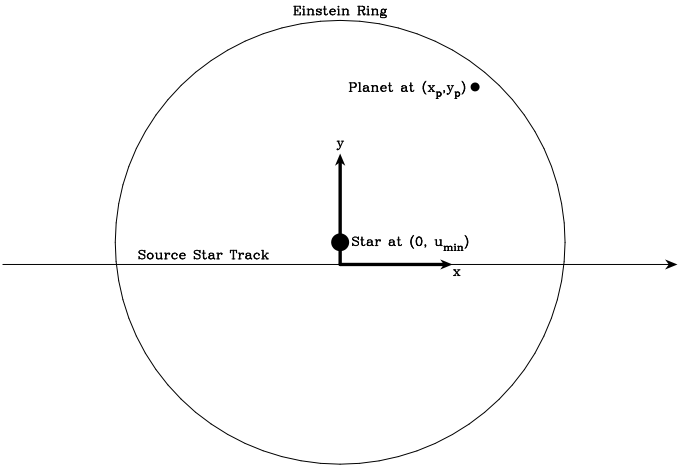}}
  \caption{Coordinate system used to analyse light curves in terms of planets 
  orbiting a lens star. The coordinates scales are in units of the Einstein 
  radius $R_{\rm E}$. }
\label{coord}
\end{figure}

The general procedure that was followed for planetary modeling was to 
calculate the $\chi^2$ values of the data for a broad range of planetary 
models to find  the model with the smallest $\chi^2$. The light curve for 
any particular planetary model was calculated using the inverse ray shooting 
technique described above. Unless otherwise stated, the computations for all 
events corresponded to the following assumed values of the lens and source 
star parameters: $M_L=0.3M_{\odot}$, $\dol$=6 kpc, $\dls$=2 kpc 
and $R_S=R_{\odot}$. These values correspond to $\dos=8$ kpc 
and $R_{\rm E}$ = 1.9 AU. The coordinate system used for the computations 
is depicted in Fig.~\ref{coord}.

For the planetary modeling, the ratio of planet-to-lens mass $\epsilon$ 
was initially allowed to vary from $10^{-7}$ to $10^{-3}$ in 33 approximately 
equally logarithmically spaced steps for each event. Similarly, the projected 
coordinates of a planet $x_{\rm p}$ and $y_{\rm p}$ at the time of peak 
magnification were allowed to vary from $-2R_{\rm E}$ to $+2R_{\rm E}$ 
in 129 equally spaced steps. Thus a total of 549,153 planetary configurations 
was initially trialed for each event. The trialing was done on a 
super-cluster computer described by Rattenbury et al. (2002). For each 
trial, the $\chi^2$ was calculated over the time interval $-t_{\rm E}$ 
to $+t_{\rm E}$. This corresponds to the source star traversing the diameter 
of the Einstein ring. Maps of $\chi^2$ over the lens plane for the 33 
planetary mass fractions were examined to determine approximate positions 
and masses of possible planets. Further minimization, described below, was 
subsequently carried out to determine the most likely planetary models.   
 
\subsection{MACHO~98--BLG--35}

\begin{figure}
 \epsfxsize=\hsize\epsfbox{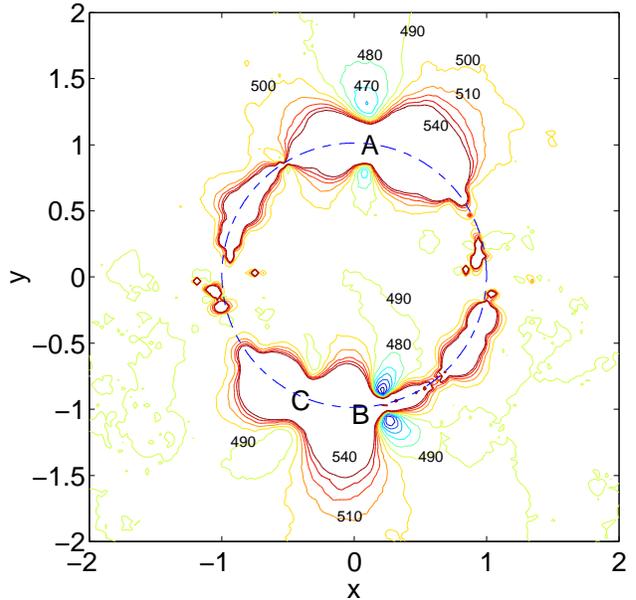}
 \caption{Map of $\chi$-squares of the data for MACHO~98--BLG--35 by the 
 MOA, MPS and PLANET groups for an Earth-mass planet ($\epsilon = 10^{-5}$) 
 orbiting the lens star at projected radii up to $2R_{\rm E}$. The coordinate 
 system is the same as in Fig.~\ref{coord}. There are three minima interior 
 to the ring, and three corresponding minima exterior to the ring, labeled 
 A, B and C respectively. These are the possible positions of terrestrial-mass 
 planets orbiting the lens star at the time of the microlensing event. Planet 
 positions of higher likelihood were determined by allowing the planet-to-lens 
 mass fraction to vary from $10^{-7}$ to $10^{-3}$ and locating deeper 
 $\chi^2$ minima.}
\label{chimap}
\end{figure}

Planetary modeling of MACHO~98--BLG--35 was carried out using the combined 
dataset from MOA, MPS and PLANET. A typical $\chi^2$ map is shown in 
Fig.~\ref{chimap}. This corresponds to an Earth-mass planet 
($\epsilon = 10^{-5}$). The minima on these maps indicate possible 
positions of planets at the time of the microlensing event. Maps for 
heavier (lighter) planets are similar, with the minima displaced further 
from (closer to) the Einstein ring. The Einstein ring is generally depicted 
clearly on these maps, either as a region where planets are strongly excluded, 
or where they may be present. The maps also depict a degeneracy that exists 
in the microlensing method, namely that planets at projected radii $a$ 
and $1/a$ (expressed in units of the Einstein radius) yield similar light 
curves, and hence are indistinguishable in this method 
(Griest \& Safizadeh 1998).                

The $\chi^2$ map shown in Fig.~\ref{chimap} for MACHO~98--BLG--35 
shows six possible 
positions for planets orbiting the lens star, two at position A, two at 
position B, and two at position C, where the doubling is caused by the 
degeneracy noted above. We have performed model fitting for each combination 
of one planet, two planet, and three planet configurations. For each model we 
carried out a $\chi^2$ minimization allowing all parameters to vary over 
small ranges near their initial values. The total number of free parameters
was 8 for one-planet models, 11 for two-planet models, and 14 for the 
three-planet models.
The minimization was achieved using the Simplex procedure. The results 
are shown in Table 3.

Among the single planet models, model A gives the best improvement over
the single lens with a renormalized $\Delta\chi^2=63.2$. Model B+C is the most 
favoured of the two planet models, although the $\chi^2$ value is not a 
significant improvement over model A. The improvement in $\chi^2$ for the
triple planet model A+B+C is also not significant and as such there is no
need to introduce a third planet. Models A and B+C appear as the most
favourable solutions for the configuration of the planetary system. While
it may seem that there is no need to introduce a two planet model, we
consider model B+C as an alternative because it is physically distinct from
the single planet model A. The ambiguity arises here because the
intensive coverage and image subtraction photometry of the event, shown
in Fig. 7, did not cover the entire full-width at half-maximum (FWHM) of
the peak. Full coverage of the FWHM would reduce the ambiguities
(Rattenbury et al. 2002).

\begin{figure*}
  \centerline{\psfig{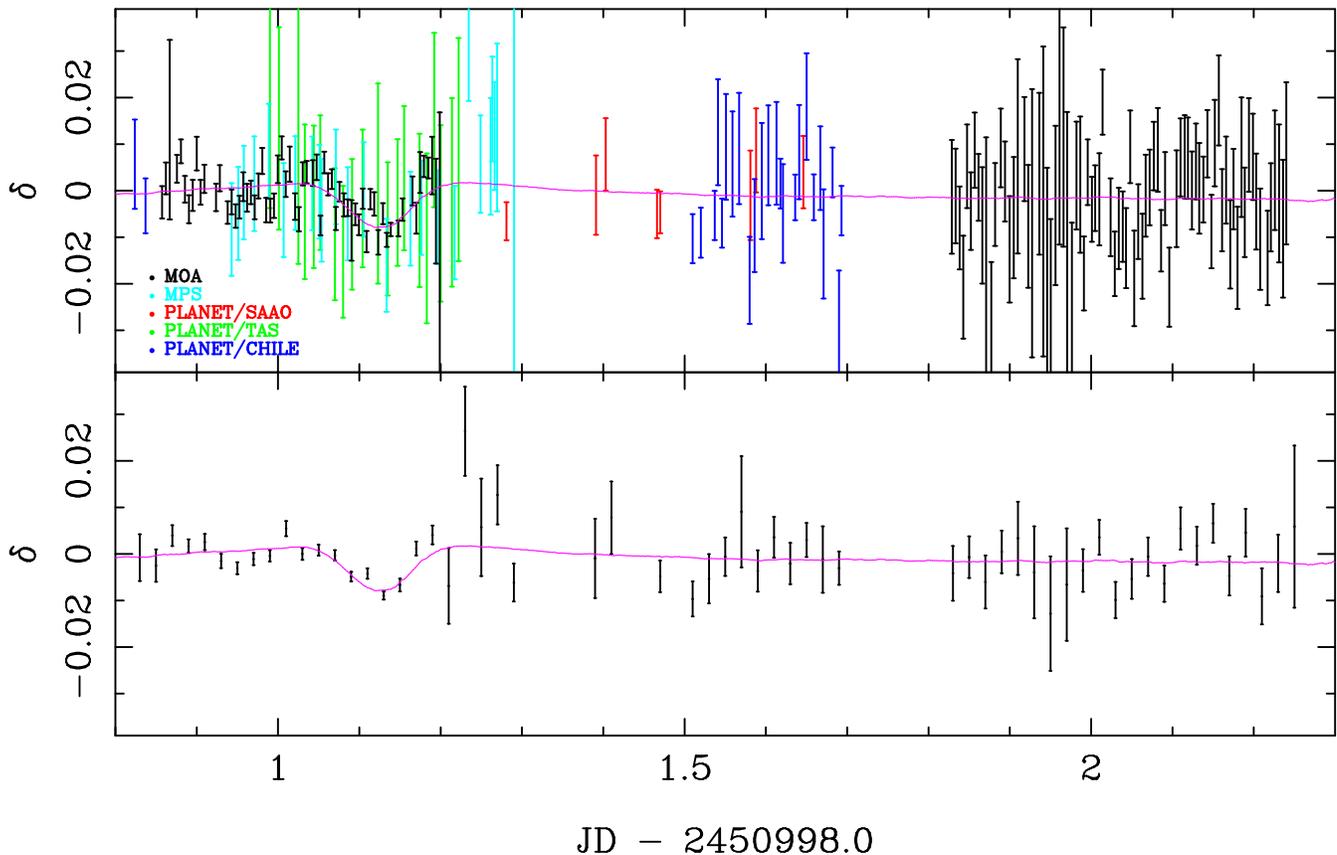}}
 \caption{Light curve for the single planet model "A" of event 
MACHO~98--BLG--35 together with the data of the MOA, MPS and PLANET groups. 
The top figure shows all the data while the lower figure shows the data 
binned and weight averaged on 0.02 day intervals. The quantity $\delta$ 
is the fractional deviation of the fit from the lensing star component of 
the star$+$planet system. The dip in the
light curve at around day 1.1 is seen in the MOA, MPS and PLANET data.}
 \label{planet1}
\end{figure*}

\begin{figure*}
  \centerline{\psfig{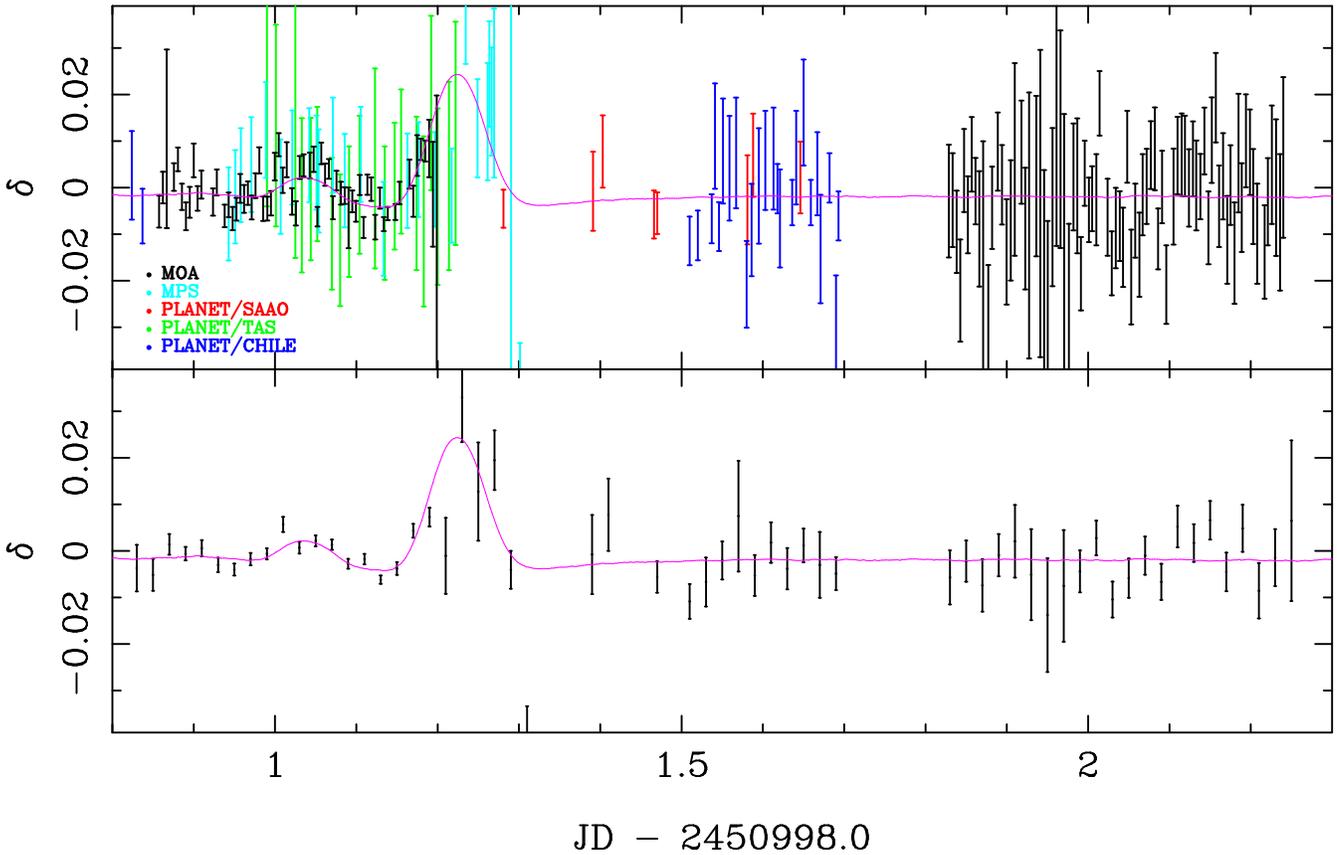}}
  \caption{Same as Fig.~\ref{planet1} but for the two planet model B+C.}
 \label{planet2}
\end{figure*}

We have drawn the light curve for model A in Fig.~\ref{planet1} and for 
model B+C in Fig.~\ref{planet2} together with the observational data. The
data are plotted in terms of the fractional deviation from the single lens
model due to the stellar component defined as
\begin{equation}
\delta = \frac{A_{\rm SP}(\tmax, \that, \umin, x_1, y_1, \epsilon_1, \cdots) - A_{\rm S}(\tmax, \that, \umin)}{A_{\rm S}(\tmax, \that, \umin)}
\end{equation}
Here $A_{\rm S}$ is the amplification due to a single lens given by 
Eqns.~2--3. This depends only on the three single lens parameters 
$\tmax$, $\that$, $\umin$. The amplification due to the star+planet system, 
$A_{\rm SP}$, depends upon the single lens parameters together with
the positions and mass ratios of the component planets. It should be
noted that the three parameters comprising the single lens component are
fitted separately for each model. This affects the shape of the profile
of the fractional deviation light curve.

\begin{figure}
  \centerline{\psfig{figure=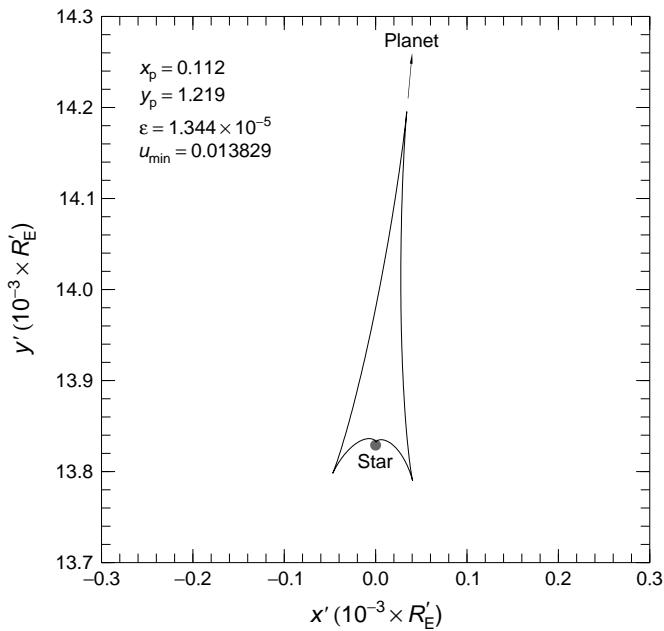,width=9cm}}
  \caption{Model A of event MACHO~98--BLG--35 viewed in the source plane. 
  The "stealth-bomber" shaped line is the "caustic" or locus of points where 
  a point source would be infinitely magnified by the lens system. The 
  source star moves horizontally left-to-right through the origin. The 
  coordinate units in the source plane are projected Einstein radii, i.e. 
  $R'_{\rm E}=(\dos/\dol)\times R_{\rm E}$. For any source radius 
  $\leq 6R_{\odot}$, the source star does not intersect the caustic.
  Rhie \& Bennett (1996), Rhie (1999), and Rhie et al (1999) discuss further
  examples of events with similar caustic geometries, and also the origin of
  the stealth-bomber terminology.}
  \label{caustic1}
\end{figure}

In Fig.~\ref{caustic1}, we depict an additional view of model A, this time in 
the source plane. This shows the contour of infinite magnification for the 
event (i.e. the ``caustic''). It is seen that the source star does not pass
between the star and the planet. Hence the perturbation is negative as
can be seen in Fig.~\ref{planet1}. It is also seen that the source star does 
not intersect the caustic at any time.

Rhie et al. (2000) estimated a radius of 1--3$R_{\odot}$ for the source star 
in this event. We have repeated the above $\chi^2$ minimization process
for model A using a source radius of 0.004$\re$($\approx2R_{\odot}$). 
We obtain best fit values for the planetary parameters of 
$\epsilon=1.3\times10^{-5}$ at position $(0.08, 1.17)$ with 
$\Delta\chi^2=36.1$. The effect of increasing the source size serves to 
partially wash out the fine structure in the microlensing light curve near 
the peak. The improvement in $\chi^2$ is then not as good as those for
smaller source radii. These results indicate a likely source star
radius $\la2R_{\odot}$.

%%%

\begin{table*}
\centering
\caption{Parameters and statistics of the best planetary fits to the combined 
data of the MOA, MPS and PLANET groups from $-t_{\rm E}$ to $+t_{\rm E}$ for 
MACHO~98--BLG--35. Model O is the best fit without a planet. The 
$\Delta\chi^2$ values are the normalized values with respect to model O. The 
coordinate 
system is as depicted in Fig.~\ref{coord}. The planetary mass fraction, 
$\epsilon$, is in units of $10^{-5}$. For each set of planetary coordinates, 
an alternative set is possible with a planet of the same mass at the inverse 
radius. For model A, for example, the conjugate coordinates are 
(0.07,0.82,1.3).} 
\label{models}
\begin{tabular}{@{}lccccccccccccc}

Model &
\multicolumn{3}{c}{Planet A} &
\multicolumn{3}{c}{Planet B} &
\multicolumn{3}{c}{Planet C} &
$t_{\rm max}$ & $t_{\rm E}$ & $u_{\rm min}$ & $\Delta\chi^2$\\

&
$x_{\rm p}$ & $y_{\rm p}$ & $\epsilon$ &
$x_{\rm p}$ & $y_{\rm p}$ & $\epsilon$ &
$x_{\rm p}$ & $y_{\rm p}$ & $\epsilon$ &\\[10pt]

O     &  -   &  -   &  -  &   -   &  -   &  -  &   -   &  -    &  -   
& 999.1494 & 20.37 & 0.013829 & 0\\
A     & 0.11 & 1.22 & 1.3 &   -   &  -   &  -  &   -   &  -    &  -   
& 999.1506 & 20.32 & 0.013829 & 60.0\\
B     &  -   &  -   &  -  & 0.30 & $-$1.11 & 2.8 &   -   &  -    &  -   
& 999.1499 & 20.31 & 0.013964 & 34.5\\
C     &  -   &  -   &  -  &   -   &  -   &  -  & $-$0.37 & $-$0.86 & 0.17 
& 999.1493 & 20.33 & 0.013898 & 13.7\\
A+B   & 0.16 & 1.25 & 0.79& 0.35 & $-$1.14 & 2.8 &   -   &  -    &  -   
& 999.1499 & 20.27 & 0.013837 & 41.7\\
B+C   &  -   &  -   &  -  & 0.30 & $-$1.12 & 2.6 & $-$0.34 & $-$0.86 & 0.19 
& 999.1500 & 20.33 & 0.013978 & 57.5\\
A+C   & 0.15 & 1.21 & 0.99&  -   &   -   &  -  & $-$0.33 & $-$0.84 & 0.18 
& 999.1508 & 20.30 & 0.013857 & 47.2\\
A+B+C & 0.19 & 1.28 & 0.30& 0.34 & $-$1.15 & 2.9 & $-$0.35 & $-$0.87 & 0.17 
& 999.1490 & 20.34 & 0.013847 & 56.1\\

\end{tabular}
\label{table3}
\end{table*}

Model B may be seen to correspond to the one previously found by Rhie et al. 
(2000) on the basis of DoPHOT analyses of the MPS and MOA datasets. The 
inclusion of the PLANET dataset has served to lower the planetary 
mass-fraction of this model from $7.0\times 10^{-5}$ 
to $2.8\times 10^{-5}$.  Model B+C may be considered a generalization of 
the original model of Rhie et al. (2000) to which a second, nearby, 
very-low-mass planet has been added. The inclusion of a second planet 
as in model B+C gives a significant decrease in $\chi^2$ over the 
single planet model B with $\Delta\chi^2=30.0$. While model A is the
simplest, we have included model B+C to illustrate the potential of
microlensing to map multi-planet systems.. 

Thus far the discussion has been in terms of $\Delta\chi^2$
values because these are both convenient for comparing different
models, and because they are less sensitive than absolute $\chi^2$ values
to the uncertainties in the measurements.
Visual inspection of Figs.~\ref{lc98} and \ref{planet1} shows that 
model O is not a good fit to the data, and
that model A is significantly better. Assuming that model A is actually
correct, a likelihood for model O may be estimated by forcing the $\chi^2$
of model A to be the same as the number of degrees of freedom by
renormalizing all the errors by a constant factor.\footnote{
A similar procedure was adopted by Albrow et al. (2001).}
This yields $\chi^2=358.0$ with 298 degrees of freedom for model O 
which corresponds to a deviation of $2.5\sigma$ for which the 
probability is $<1$\%.
A more satisfactory procedure for interpreting MACHO~98-BLG-35 would be to 
analyze all the datasets using difference imaging, determine all the errors 
self-consistently, and redo the planetary modelling.

For the two planetary configuration models that we favour here, model A and
model B+C, the renormalized $\Delta\chi^2$ was around 60 and the number of 
degrees of freedom were around 290. Gaudi et al (2002) subsequently proposed
setting an alternative detection threshold for low-mass planets,
namely $\Delta\chi^2>60$ irrespective of the number of degrees of freedom.
Howver, as the chance probability depends on the number of degrees of freedom,
we advocate consideration of this factor. Gaudi et al also noted that 
heavy planets should be detected before light planets unless the planetary
mass function is steep. We note that, in the absence of observational
information on the planetary mass function, a steep function cannot be excluded.
Gaudi et al analysed PLANET data for several microlensing events, 
including MACHO~98--BLG--35, and reported no evidence for planets. Our 
representation of their data in Figs.~12 and 13 is consistent with this.
However, it is clear from Figs.~12 and 13 that the PLANET data,
derived using profile fitting photometry, are not in conflict with the
MOA data, derived using the more precise image subtraction photometry. As
such the PLANET data do not rule out the planetary models we considered in
our study.

Exclusion regions for giant planets orbiting the lens star of 
MACHO~98--BLG--35 were also computed using the inverse ray shooting 
technique. The $\chi^2$ maps for one-planet models for mass ratios of 
($\epsilon=2.8\times 10^{-3}$), ($\epsilon=8.5\times 10^{-4}$) and 
($\epsilon=1.3\times10^{-4})$ were computed. These correspond to 
Jupiter, Saturn, and Uranus mass planets orbiting a one third solar 
mass star. The contours were found where $\chi^2$ exceeds its value 
for the single lens model by 90. This value was chosen, under the 
assumption that any excluded model should have $\sim20$ or more 
consecutive measurements deviating systematically by $(1-2)\sigma$ or 
more from it. This threshold would appear conservatice when compared with 
the $\Delta\chi^2$ values for the planetary models. The higher value was
chosen to allow for the fact that the other parameters 
$t_{\rm max}$, $t_{\rm E}$, and $u_{\rm min}$ were not allowed to float 
but were fixed at the single lens model values in this computation. The 
exclusion regions so obtained are shown in Fig.~\ref{exclude1}. It is 
clear that a large region surrounding the lens star of MACHO~98--BLG--35 
is devoid of gas-giants like Jupiter or Saturn. Jovian planets with 
projected radii $\sim$0.2--15 $R_{\rm E}$ or $\sim$0.4--30 AU are excluded. 
A similar result was found previously by Rhie et al. (2000), but the 
inclusion of further data and the use of subtraction photometry have 
served to enlarge the exclusions regions.

\subsection{MACHO~99--LMC--2}

The microlens event MACHO~99-LMC-2 was unusual in that it is the only high 
magnification event observed to date that occurred towards the Magellanic 
Clouds. It was found independently by both the MACHO and the OGLE groups. 
It was included in the observing program of the MOA group because it afforded 
an opportunity to search for a planet in an external galaxy, i.e. for an 
extra-galactic planet, under the assumption that the event is an example 
of the ``self-lensing'' process of Sahu (1994). In this process, a 
foreground star in the LMC lenses a background star in the LMC. We note, 
however, that the question of the most likely location of lenses observed 
towards the LMC has not yet been settled (see, for example, 
Alves \& Nelson 2000, Alcock et al. 2001). 

The typical value of the Einstein radius for self-lensing is the same as 
for galactic bulge events, i.e. $\sim2$ AU. This follows from Eqn.~1
assuming a typical lens mass $\sim0.3 M_{\odot}$ and values of order 
48 kpc, 2 kpc and 50 kpc for $\dol$, $\dls$ and $\dos$. Consequently, 
high magnification events observed towards the LMC offer similar 
prospects for planet detection as they do towards the galactic bulge.

The data for MACHO~99-LMC-2 which are displayed in Fig.~\ref{lc99} 
and Table.~2 show no clear evidence for a deviation from the light 
curve of a single lens. Consequently, only exclusion regions for 
giant planets surrounding the lens were computed. This was done 
with the procedure used above for MACHO~98--BLG--35. The exclusion 
regions so obtained are displayed in Fig.~\ref{exclude2}. Jovian 
planets with projected radii $\sim$0.4--10 AU are excluded. These 
results are, to our knowledge, the first limits placed on planetary
companions to an extra-galactic star.

\subsection{OGLE~00-BUL-12}

The data on OGLE~00-BUL-12 in Fig.~\ref{lcogle} and Table~2 show possible 
evidence of a deviation from the behaviour of a single lens. Consequently, 
they were subjected to the same analysis as that accorded to 
MACHO~98--BLG--35. The 
initial search with the super-cluster computer yielded several possible 
planetary models, but none of these were statistically significant 
improvements over the single lens fit. However, one of them showed 
interesting behaviour, and this is reproduced below in Figs.~\ref{mercury} 
and \ref{caustic3} to illustrate the potential of the high magnification 
technique.  

The light curve shown in Fig.~\ref{mercury} shows the characteristic 
double-spiked behaviour of a caustic crossing. This is illustrated in 
Fig.~\ref{caustic3}
which shows the source plane view of the model. Here a source radius of 
$1R_{\odot}$ has been assumed. The combination of high magnification and 
caustic crossing occurring simultaneously in one event leads to enhanced 
sensitivity for planet detection. The planetary mass-fraction for the light 
curve in Fig.~\ref{mercury} ($\epsilon = 0.055 \times 10^{-5}$) 
corresponds to a planet 
of mass similar to that of Mercury orbiting a one-third solar mass star.
The improvement over the single lens fit corresponds to $\Delta\chi^2=13.7$.
We note that further data were obtained during the peak of this event 
(Sackett 2001). These could be analysed by the image subtraction technique
and included in a future analysis of the event, thus reducing the uncertainties
in the present analysis.

Given the limited coverage of this event and the low statistical significance,
we do not present this as a planetary detection. 
However, the analysis
raises interesting possibilities. The striking feature of this
event is its very high peak magnification of about 160. Light curves such as
that shown in Fig.~\ref{mercury} for a Mercury mass planet should apply
to other events with similarly high magnifications. It can be seen that such
planets can produce deviations from the single lens by about 1\%. Such
deviations should be detectable by a network of 1-m class telescopes
providing continuous and complete coverage of the peaks of high 
magnification events.

Exclusion regions for gas-giant planets orbiting the lens of OGLE~00-BUL-12 
were also computed. These are shown in Fig.~\ref{exclude3}. Jovian planets 
with projected radii $\sim$0.2--30 AU are excluded.
  
\begin{figure*}
  %\centerline{\psfig{figure=exclude1.ps,width=17.7cm}}
  \caption{Exclusion regions for giant planets similar to Jupiter, Saturn and Uranus orbiting the lens star of event MACHO~98--BLG--35. The coordinate system is the same as in Fig.~\ref{coord}. The reflection symmetry about the Einstein ring that is discussed in \S5.1 is evident.}
\label{exclude1}
\end{figure*}

\section{Discussion and Conclusion}

The original proposal of Griest \& Safizadeh (1998) to study extra-solar 
planets in gravitational microlensing events of high magnification has 
received support from the present work. It has been demonstrated that 
terrestrial-mass planets at orbital radii $\sim2$ AU may be detected 
with 1-m class telescopes. The essential requirement is the relentless 
observation of the peaks of high magnification events with coordinated 
telescopes encircling the globe, a relatively straightforward task. The 
dense stellar fields in which microlensing is necessarily observed require 
special photometric techniques to take account of the blending of stellar 
images. Difference imaging analysis appears to be well suited to this task.   

If a concerted effort was made by the microlensing community to detect and 
monitor relentlessly future events of high magnification, it could reasonably 
be anticipated that a first approximate measurement of the abundance of 
terrestrial planets could be obtained in a few years. If $\sim10$ high
magnification events can be detected per year, a rough measurement or upper
limit on the abundance of Earth-mass planets orbiting $\sim$0.3$M_\odot$ stars
at projected radii $\sim$1.5--2.5 AU could be obtained in a few years
(Bond et al 2002). As demonstrated here, 
one might also obtain new information on planets as light as Mercury, on 
extra-galactic planets, and on gas-giants. Also, multi-terrestrial-planet 
systems should be able to be mapped in some cases (Rattenbury et al 2002), 
rather like the multi-gas-giant systems presently being mapped by the 
radial velocity community (Butler et al. 1999). Ultimately, a dedicated 
wide-angle space-borne telescope
such as that planned for the proposed GEST mission would
provide excellent sensitivity towards all solar-system analogues except
Mercury and Pluto (Bennett et al 2002). 

It appears likely that none of the three lens objects studied here 
contains a Jovian planet. This is consistent with previous measurements 
by Marcy \& Butler (1998) and of Albrow et al. (2001) of other systems. 
The dearth of Jupiter-like planets enhances the sensitivity of microlensing 
studies of terrestrial planets, merely through the lack of ``background'' 
signals from gas-giants.

For all three events further data by microlensing groups exist that could 
be analysed using image subtraction and incorporated in the planetary 
modeling. This could be expected to produce results of greater precision, 
and to diminish ambiguities that presently exist in the planetary modeling. 

The assumptions made in the present analysis on the masses, sizes and 
distances of the lens and source stars may be able to be avoided in the 
future. When the Next Generation Space Telescope comes into operation, 
and as the lens and source stars in today's microlensing events begin to 
diverge, it should be possible to measure the lens and source stellar 
parameters in these events by elementary photometry (Rattenbury et al 2002). 
This being the case, 
the planetary modeling would be entirely free of undetermined 
parameters\footnote{We have examined a 50-year old image of the region 
containing MACHO~98--BLG--35, but unfortunately the lens and source stars 
were not identifiable.} 
              
\begin{figure}
  %\centerline{\psfig{figure=exclude2.ps,width=9cm}}
  \caption{Exclusion regions for gas-giants similar to Jupiter, Saturn and 
  Uranus orbiting the lens star of event MACHO~99-LMC-2. The coordinate 
  system is the same as in Fig.~\ref{coord}.}
\label{exclude2}
\end{figure}

\begin{figure}
  \centerline{\psfig{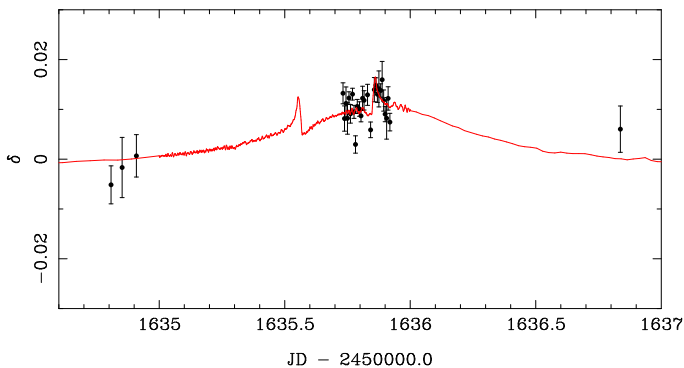}}
  \caption{Light curve around the peak of OGLE~00-BUL-12 for the best low 
  mass planet model fit together with the photometry measurements derived 
  from image subtraction. The jaggedness throughout the model curve
  is small amplitude computational noise.}
  \label{mercury}
\end{figure}
\begin{figure} 
  \centerline{\psfig{figure=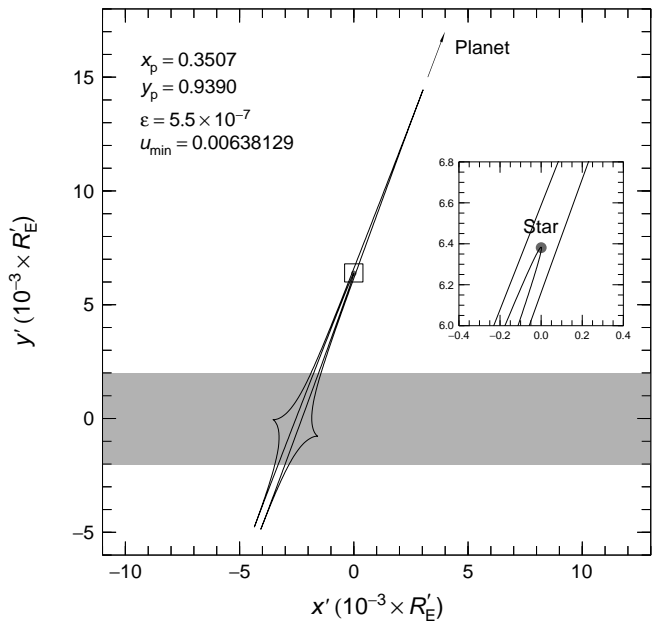,width=9cm}}
  \caption{Planetary model of event OGLE~00-BUL-12 that includes an 
  arrow-like caustic. The axes are as in Fig.~\ref{caustic1}. The source 
  star moves horizontally left-to-right through the  origin, and thus 
  intersects the caustic whatever its radius. The shaded region corresponds 
  to the track of a source star of radius of $1R_{\odot}$. Similar arrow-like 
  caustics appear in Wambsganss (1997).}
\label{caustic3}
\end{figure}

We conclude that coordinated observations using 1-m class telescopes of 
gravitational microlensing events of high magnification are capable of 
making valuable contributions to the study of planets orbiting other stars.  

\section*{Acknowledgments}

We are very grateful to the OGLE collaboration for kindly making their images
available to us. In particular we thank Andrzej Udalski and Karol Zebrun for
providing images. We are also indebted to Dave Bennett and Sun Rhie for 
introducing us to the high magnification technique, Joachim Wambsganss for 
general discussion on microlensing, and Peter Dobcsanyi for introducing us 
to super-cluster computing. The Marsden Fund of New Zealand and the Ministry 
of Education, Science, Sports and Culture of Japan are thanked for the 
financial support that made this work possible.

\begin{figure*}
  %\centerline{\psfig{figure=exclude3.ps,width=17.7cm}}
\caption{Exclusion regions for gas-giants similar to Jupiter, Saturn and 
Uranus orbiting the lens star of event OGLE~00-BUL-12. The coordinate 
system is the same as in Fig.~9. The outer extremities of the exclusion 
region for a Jupiter analogue are given by the reflection symmetry 
$a \sim1/a$.}
\label{exclude3}
\end{figure*}

\label{lastpage}


\begin{thebibliography}{}
\bibitem{} Abe, F., et al., 1997, in ``Variable Stars and the Astrophysical 
Returns of Microlensing Surveys'', eds. R.Ferlet, J. Maillard, and B. Raban, 
Editions Frontieres, France, p. 75.
\bibitem{} Albrow, M.D., et al., 2001, ApJ, 556, L113
\bibitem{} Alard, C., Lupton R., 1998, ApJ, 503, 325
\bibitem{} Alard, C., 1999, A\&A, 342, 10
\bibitem{} Alcock, C., et al., 2001, ApJ, 552, 582
\bibitem{} Alves, D., Nelson C., 2000, ApJ, 542, 789
\bibitem{} Bennett, D.P. et al., 2002, in AAS 199$^{\rm th}$ meeting, 
Washington DC
\bibitem{} Bennett, D.P., et al., 1999, Nat., 402, 57
\bibitem{} Bennett, D.P., Rhie S.H., 1996, ApJ, 472, 660
\bibitem{} Bolatto, A.D., Falco, E.E., 1994, ApJ, 436, 112
\bibitem{} Bond, I.A., et al., 2002, submitted to MNRAS
\bibitem{} Butler, R.P., Marcy, G. W., Fischer, D. A., Brown, T. M.,
 Contos, A. R., Korzennik, S. G., Nisenson, P., Noyes, R. W., 
 1999, ApJ, 526, 916
\bibitem{} Charbonneau, D., Brown, T. M., Latham, D. W., Mayor, M., 
 2000, ApJ 529, L45 
\bibitem{} Gaudi, B.S. et al., 2002, astro-ph/0104100, ApJ, in press
\bibitem{} Gould, A., Loeb A.,1992, ApJ, 396, 104
\bibitem{} Griest, K., Safizadeh N., 1998, ApJ, 500, 37
\bibitem{} Henry, G.W., Marcy, G.W., Butler, P., Vogt, S., 2000, ApJ, 529, L41
\bibitem{} Jha, S., Charbonneau, D., Garnavich, P. M., Sullivan, D. J.,
 Sullivan, T., Brown, T. M., Tonry, J. L., 2000, ApJ, 540, L45
\bibitem{} Liebes, S., 1964, Phys. Rev., 133, B835
\bibitem{} Mao, S., Paczynski B., 1991, ApJ, 374, L37
\bibitem{} Marcy, G.W., Butler R.P., 1998, ARA\&A, 36, 57 
\bibitem{} Mayor, M., Queloz D., 1995, Nature, 378, 355
\bibitem{} Paczynski, B., 1986, ApJ, 304, 1
\bibitem{} Perryman, M.A.C., 2000, Rep. Prog. Phys., 63, 1209
\bibitem{} Rattenbury, N.J., Bond, I.A., Skuljan, J., Yock, P.C.M., 2002,
MNRAS, submitted
\bibitem{} Rhie, S.H., 1999, in Gravitational Lensing: Recent Progress and 
Future Goals, 25-30 July 1999, Boston University
\bibitem{} Rhie, S.H., Becker, A. C., Bennett, D. P., Fragile, P. C., 
 Johnson, B. R., King, L. J., Peterson, B. A., Quinn, J, 1999, ApJ, 522, 1037 
\bibitem{} Rhie, S.H., Bennett, D.P., 1996, Nucl. Phy. Supp., 51B, 86
\bibitem{} Rhie, S.H. et al., 2000, ApJ, 533, 378
\bibitem{} Sackett, P.D., 2001, in Planetary Systems in the Universe:
Observations, Formation, and Evolution, ASP Conf. Series, eds A.J. Penny, 
P. Artymowicz, A.-M. Lagrange, and S.S. Russell, in press
\bibitem{} Sahu, K.C., 1994, Nature, 370, 275  
\bibitem{} Schechter, P., Mateo M., Saha A., 1993, PASP, 105, 1342
\bibitem{} Udalski, A., Kubiak M., Szymanski M., 1997, AA 47, 319
\bibitem{} Wambsganss, J., 1997, MNRAS, 284, 172
\bibitem{} Yanagisawa, T., et al., 2000, Exp. Astron., 10, 519
\end{thebibliography}
\end{document}